\documentclass[aps,pre,twocolumn,superscriptaddress]{revtex4-1}
\usepackage{graphicx}
\usepackage[rgb,dvipsnames]{xcolor}
\usepackage{amsmath}
\usepackage{amsthm, amssymb}
\usepackage{epsfig}
\usepackage{hyperref}
\usepackage{bm}

\usepackage{natbib}

\usepackage[normalem]{ulem}

\usepackage{placeins}
\usepackage[symbol]{footmisc}

\usepackage{tikz}
\usetikzlibrary{trees}

\begin{document}

\title{Precision of morphogen-driven tissue patterning during development is enhanced through contact-mediated cellular interactions}

\author{Chandrashekar Kuyyamudi}
\affiliation{The Institute of Mathematical Sciences, CIT Campus, Taramani, Chennai 600113, India}
\affiliation{Homi Bhabha National Institute, Anushaktinagar, Mumbai 400 094, India}
\author{Shakti N. Menon}
\affiliation{The Institute of Mathematical Sciences, CIT Campus, Taramani, Chennai 600113, India}
\author{Sitabhra Sinha}
\affiliation{The Institute of Mathematical Sciences, CIT Campus, Taramani, Chennai 600113, India}
\affiliation{Homi Bhabha National Institute, Anushaktinagar, Mumbai 400 094, India}
\date{\today}
\begin{abstract}
Embryonic development involves pattern formation characterized by the emergence of
spatially localized domains
characterized by distinct cell fates resulting
from differential gene expression.
The boundaries demarcating these domains are precise and consistent within a species
despite stochastic fluctuations in the morphogen molecular concentration that provides positional information to the cells,
as well as, the intrinsic noise in molecular processes that interpret this information to guide
fate determination.
We show that local interactions between physically adjacent cells mediated by
receptor-ligand binding utilizes the asymmetry between the fate-determining genes
to yield a switch-like response to the global signal provided by the morphogen.
This results in robust developmental outcomes with a consistent identity of the gene that is dominantly expressed
at each cellular location, thereby substantially reducing the uncertainty in
the location of the boundary between distinct fates.
\end{abstract}


\maketitle

\begin{figure}[tbp!]
\centering
\includegraphics[width=0.99\columnwidth]{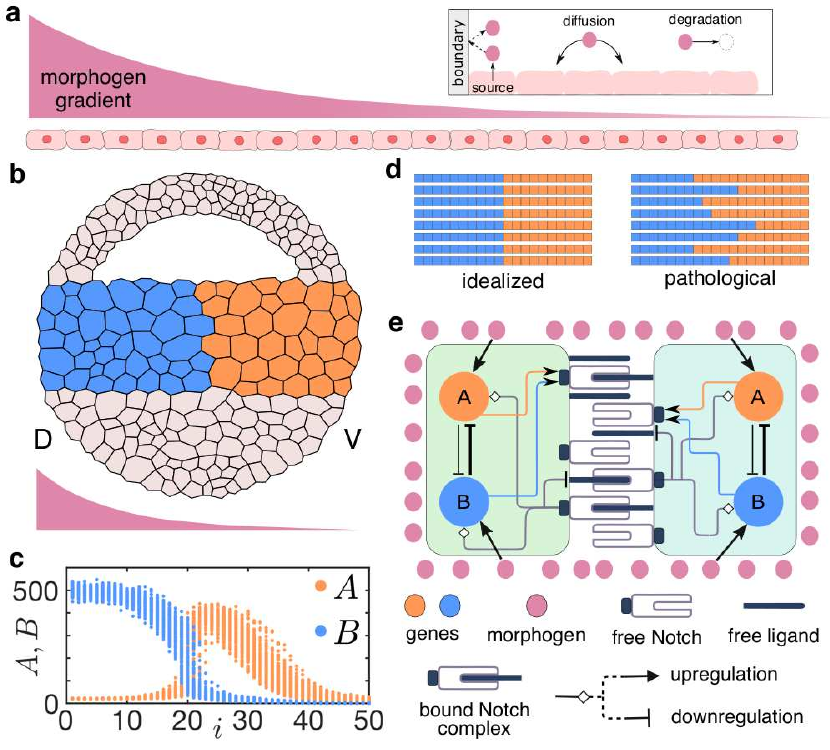}
\caption{\textbf{Cell fate determination through a morphogen concentration gradient
needs to be robust against stochastic fluctuations.}
(a)~A morphogen gradient across a cellular array results from the processes  (shown in
the inset) of
secretion of molecules from a source located at the boundary of the domain, their diffusion across space and degradation
over time such that the decay rate is linearly proportional to its concentration. (b)~Schematic
representation of a \textit{Xenopus} embryo where the differentiation of the cells of the
mesoderm into dorsal and ventral fates (represented by blue and orange, respectively)
is guided by the concentration gradient of the morphogen \textit{activin} between the
dorsal (D) and ventral (V) ends (displayed below the embryo).
(c)~The steady state expression of patterning genes $A,B$  across a $1$-dimensional array
comprising $N$ cells, with the indices of the
cells indicated by $i =1, \ldots, N (=50)$, subject to a noisy morphogen gradient in
the absence of interaction between the cells.
Results of $300$ different realizations are shown.
(d)~While in the absence of noise the boundary separating the regions with the two different
fates corresponding to $B>A$ (blue) and $A>B$ (orange) is expected to occur at the
same position across all realizations (the idealized situation shown at left), fluctuations
in the morphogen concentration and gene expression dynamics results in variations across
realizations (shown at right) if fate determination occurs only on the basis of
positional information provided by the morphogen gradient.
(e)~Interactions between neighboring cells mediated by Notch-Delta signaling pathway
(shown here schematically) can aid in the robust determination of fate boundaries in
the presence of noise. Genes $A$ and $B$ comprising the morphogen interpretation module
affect the expression of genes coding for Notch receptors. The Notch Intracellular Domain (NICD), released from the bound Notch complex that results from the \textit{trans}-activation of Notch receptors, in turn up- or downregulates the expression of $A$ and $B$ (depending on the type
of interaction).
}
\label{fig:fig1}
\end{figure}

The ubiquity of noise in the natural world makes it imperative that biological processes
are robust to it~\cite{Kitano2004,Tsimring2014}. This is particularly relevant during the development of an organism as small deviations resulting from chance events at earlier stages
can get amplified over time leading to pathological outcomes~\cite{Waddington1957,Wolpert2015}.
Indeed, embryos
exhibit a highly reproducible sequence of cellular division, differentiation and rearrangement
resulting in a physiological organization that is consistent across all individuals of a species~\cite{Sternberg2004,Lander2013,Gilbert2013}.
Morphogenesis involves pattern formation~\cite{Cross1993,Koch1994} in which cells at various locations
in a tissue
adopt distinct specialized roles (fates) via differential gene expression. This is often guided
by concentration gradients of molecules known as morphogens that emerge
via diffusion from localized sources [Fig.~\ref{fig:fig1}~(a)].
Each cell responds to the local morphogen concentration in its immediate neighborhood
and attains a fate determined by whether the  concentration lies between a specific pair
of thresholds~\cite{Wolpert1969,Wolpert1989,Gurdon2001,Sharpe2019}.
The resulting domains with different fates are characterized by sharp boundaries
whose locations are invariant for a species, e.g., that occurring between cells expressing
dorsal and ventral fates in an embryo [shown in Fig.~\ref{fig:fig1}~(b) for \textit{Xenopus}].
This is surprising as,
in order to adopt a fate consistent with its position, a cell must correctly infer
its location in the tissue from the information provided by the morphogen
concentration signal, which is very noisy due to fluctuations in the synthesis, degradation and
diffusive transport of molecules [Fig.~\ref{fig:fig1}~(a), inset]~\cite{Gurdon2001,Lander2002,Hornung2005,Kicheva2007}.
In addition, each of the steps involved in the intra-cellular response, from binding of
morphogen with surface receptors to the downstream signaling cascade terminating in
gene expression, is inherently noisy because of the underlying probabilistic processes
involving a small number of molecules involved ($\ll N_o$,
Avogadro's number)~\cite{Elowitz2002,Kaern2005,Arias2006}.

Fig.~\ref{fig:fig1}~(c) shows that, in the absence of any explicit mechanism for noise reduction,
the expression levels of a pair of patterning genes $A$ and $B$ in the cells of a model system responding to
the local morphogen concentration are subject to a high degree of variation.
The expression levels are observed to be comparable over a number of cells
such that neither gene is guaranteed to dominate and hence determine the fate,
suggesting that the cell fates are primarily decided by random chance events~\cite{Zheng2018,Guillemin2020}.
This would
result in the length of the domains comprising cells with different fates
varying considerably across realizations, which contrasts sharply with the highly reproducible spatial
pattern that is expected [Fig.~\ref{fig:fig1}~(d)]. Thus, processes that aid in reducing
variability must underlie the high level of precision in fate boundaries observed during
development~\cite{Hansen2018,Exelby2021}. Among the several candidate mechanisms that have been proposed, many
involve making the behavior of the morphogen interpretation module within each cell
more robust, e.g., incorporating the dynamics of genetic regulatory networks~\cite{Lagha2012,Chalancon2012,Perez2016,Exelby2021}.
Alternatively, consistency in cell fate decision-making can be promoted by regulating the
nature of the morphogen concentration gradient so as to reduce fluctuations in it~\cite{Hu2010,Cotterell2010,Guillemin2020}.
In general, all such mechanisms that improve the reliability of cellular decision-making based on
their spatial location can be considered to effectively pool together information gathered
from multiple measurements of the morphogen signal in the immediate neighborhood~\cite{Lander2013}.
While for a single cell, this typically involves temporal integration
of the signal, the same aim can potentially be achieved by neighboring cells sharing
information about the morphogen concentration that they each detect~\cite{Mugler2016}.
As in the developing embryo, cells in close physical proximity are known to communicate
with each other through contact-mediated signaling, such inter-cellular interactions
can be a possible mechanism through which spatial integration of the morphogen
signal can be implemented~\cite{Ellison2016,Lander2013}. One of the most widely observed examples of such interactions
is the evolutionarily conserved Notch signaling
pathway~\cite{Artavanis1999,Sprinzak2010,Sprinzak2011,Kuyyamudi2021_b}, which is triggered by Notch receptors on the surface
of a cell binding to membrane-bound proteins (e.g., Delta ligand) of a neighboring cell.
Indeed, Notch-mediated interactions are known to have a fundamental role in all metazoan development~\cite{Artavanis1999,Kopan2009}.
Although it has been suggested earlier that such contact-mediated signaling may play a role in regulating noise~\cite{Erdmann2009,Lander2011,Lander2013}, the mechanism through
which this can arise is yet to be established.

In this paper we have demonstrated that the precision of the boundary between
domains expressing different cell fates is improved considerably when cells
can communicate via Notch signaling. Specifically, we investigate the role
played by such signals in regulating the
expression of mutually inhibiting patterning genes ($A,B$)
that determine the developmental fate of a cell.
Noise, in the form of
stochastic fluctuations in the concentration of the morphogen, as
well as, in that of the signaling molecules and the expression levels of the patterning genes,
results in a high degree of variability in the fate adopted by each cell in isolation.
However, when the downstream effector ($S$) of the Notch signaling pathway is allowed to
upregulate the patterning gene that can express at a lower morphogen concentration
(assumed to be $A$)
compared to the other, we observe a remarkable decrease in the uncertainty in
the fate of a cell at a particular location in the tissue. The effectively equivalent
interaction in which the other gene ($B$) is downregulated by the signal also shows a qualitatively
similar outcome. In contrast, for interactions of the opposite type (viz., $S$ upregulating
$B$ or downregulating $A$), an increase in the sharpness of fate boundaries is seen
over a more limited region of the relevant parameter space.
Insight into the process by which the coupling counters noise is provided by the observation
that robustness requires the time-scale of the contact-induced signal to be
longer than those associated with gene expression dynamics. Our results show that
Notch signaling between cells is capable of exploiting any inherent asymmetry in the interactions between patterning genes and their response to the morphogen, yielding a highly robust
developmental outcome.

To investigate the potential role of contact-mediated interaction between cells in generating
robust spatial patterns from position-dependent cell fate determination in the presence
of stochastic fluctuations,
we consider a linear array of cells that are subject to a morphogen concentration gradient.
The source from which the morphogen molecules are secreted at a constant rate $\alpha_M$
is assumed to be located at one end of the array. The molecules, that decay after a mean lifetime
$\tau_M$, randomly disperse in a medium having diffusion coefficient $D_M$ across the array,
resulting in their concentration exhibiting fluctuations around an exponentially decaying
spatial profile. The temporally averaged
signal strength sensed
by a cell located at a distance $x$ away from the source is $M(x) = M(0) \exp(-x/\lambda_M)$,
where $\lambda_M$ is the characteristic length scale associated with the gradient.
At any instant, the magnitude of the signal governs the expression of genes comprising
the morphogen interpretation module. We choose the simplest non-trivial example of
differential gene expression leading to spatial patterning, viz.,  a module
having two genes, $A$ and $B$ [Fig.~\ref{fig:fig1}~(e)].
As is characteristic of gene circuits that respond to the concentration of an external
morphogen, the two patterning genes are
assumed to mutually repress each other, favoring the dominance of one over the other
in terms of expression levels~\cite{Gurdon2001,Ashe2006}.
The maximally expressed gene among the two within each cell decides its corresponding
fate.
For example, in the context of mesoderm differentiation in \textit{Xenopus} in the presence
of the morphogen \textit{activin}, they can be identified with the genes \textit{Goosecoid} and \textit{Brachyury}~\cite{Smith1995,Saka2007}.
Here we focus on the location of the fate boundary that demarcates regions
with high levels of expression of $A$ from those of $B$.

As mentioned above, the expression of gene $A$ occurs at relatively low values of the signal, unlike gene $B$ which needs higher concentration of the morphogen. Thus, to prevent
a homogeneous fate for the entire domain, we need to ensure that higher concentrations of the
morphogen favor the expression of $B$. This is achieved by an asymmetric mutual
repression such that $B$ inhibits $A$ more strongly than $A$ does $B$.
Contact-mediated interaction between cells is implemented by coupling the patterning gene expression dynamics of adjacent cells through Notch signaling~\cite{Artavanis1999,Kopan2009}.
Specifically, when both genes are expressed at high levels in a cell, it results in upregulation
of the gene encoding Notch, leading to an increased concentration of free receptors ($R$). This
enhances the  strength of contact-mediated interactions by increasing the
the probability of a binding event.
The trans-activation of Notch receptors upon binding to a Delta ligand of a neighboring cell
leads to a downstream effector $S$ of the resulting signaling cascade regulating the expression
of the patterning genes. Based on whether $S$ up or downregulates the expression of
gene $A$ or gene $B$, we can classify the intercellular interactions into four different types.
We report below in detail the dynamical consequences of each type of coupling.
The signaling resulting from trans-activation of Notch receptors also results in the repression
of the production of Delta ligand protein~\cite{Sprinzak2010,Barad2010}, thereby decreasing the concentration of free ligands ($D$).
%

The equations describing the stochastic dynamics of all variables ${\bm X}: \{M,A,B,R,D,S\}$ in our model have the form $d {\bm X} =
\mathcal{F}_{\bm X} dt + \mathcal{G}_{\bm X} dW$, with the stochastic component
being $\mathcal{G}_{\bm X} = \eta {\bm X}$ where $\eta$ is the strength of the noise
and $dW$ is a Wiener process~\cite{Van1992,Higham2001},
while the deterministic component $\mathcal{F}$ for the different variables of the
system are given by:
\begin{align*}
&\mathcal{F}_M  = \alpha_M\delta_{i,1} - D_M \nabla^2 M - \frac{M}{\tau_M}\,,\\
&\mathcal{F}_A = \alpha_A \mathcal{H}_h(M,K_1) \mathcal{H}^{\prime}_h(B,K_3) \Phi_A + \gamma_A \mathcal{H}_g(S,Q) - \frac{A}{\tau_A}\,,\\
&\mathcal{F}_B = \alpha_B \mathcal{H}_h (M,K_2) \mathcal{H}^{\prime}_h(A,K_4) \Phi_B + \gamma_B \mathcal{H}_g(S,Q) - \frac{B}{\tau_B}\,,\\
&\mathcal{F}_R = \beta_{R_0} + \beta_{R} \mathcal{H}_g (A,J) \mathcal{H}_g (B,J) - k_{tr}RD_{tr} - \frac{R}{\tau_R}\,,\\
&\mathcal{F}_D = \beta_{D_0} + \beta_{D} \mathcal{H}^{\prime}_g(S,K_5) - k_{tr}R_{tr}D - \frac{D}{\tau_D}\,,\\
&\mathcal{F}_S  = k_{tr}RD_{tr} - \frac{S}{\tau_S}\,,
\end{align*}
where $R_{tr}$ and $D_{tr}$ refers to the total concentrations of
receptors and ligands, respectively, in the neighboring cell(s).
The Hill functions corresponding to activation and inactivation of $X$ are described as $\mathcal{H}_{\beta}(X,C)=X^{\beta}/(C^{\beta}+X^{\beta})$ and $\mathcal{H}^{\prime}_{\beta}(X,C)=C^{\beta}/(C^{\beta}+X^{\beta})$, respectively, with $C$ as the half-saturation constant
and $\beta$ being the Hill exponent. The functions $\Phi_A, \Phi_B$ and parameters
$\gamma_A,\gamma_B$ characterize the four distinct
types of inter-cellular interactions and are defined in Table~\ref{tab:tab1}.
\begin{table}[htbp!]
\centering
\begin{tabular}{|c|c|c|c|c|}\hline
  & $\Phi_A$ & $\Phi_B$ & $\gamma_A$ & $\gamma_B$ \\\hline
$S\ \raisebox{-.2ex}{\rotatebox{90}{\scalebox{1}[2.1]{$\bot$}}}\ B$ & 1 & $Q^g/Q^g+S^g$ & 0 & 0 \\\hline
$S \longrightarrow A$ & 1 & 1 & $>$0 & 0 \\\hline
$S\ \raisebox{-.2ex}{\rotatebox{90}{\scalebox{1}[2.1]{$\bot$}}}\ A$ & $Q^g/Q^g+S^g$ & 1 & 0 & 0 \\\hline
$S \longrightarrow B$ & 1 & 1 & 0 & $>$0 \\\hline
\end{tabular}
\caption{Description of the functions and parameters defining the four different types of
inter-cellular signaling considered, based upon the nature of interaction, viz., upregulation
($\rightarrow$) or downregulation ($\dashv$~),
and the identity of the patterning gene whose expression is regulated by the
Notch downstream effector $S$, i.e., $A$ or $B$.}
\label{tab:tab1}
\end{table}


\begin{figure}[tbp!]
\includegraphics[width=0.99\columnwidth]{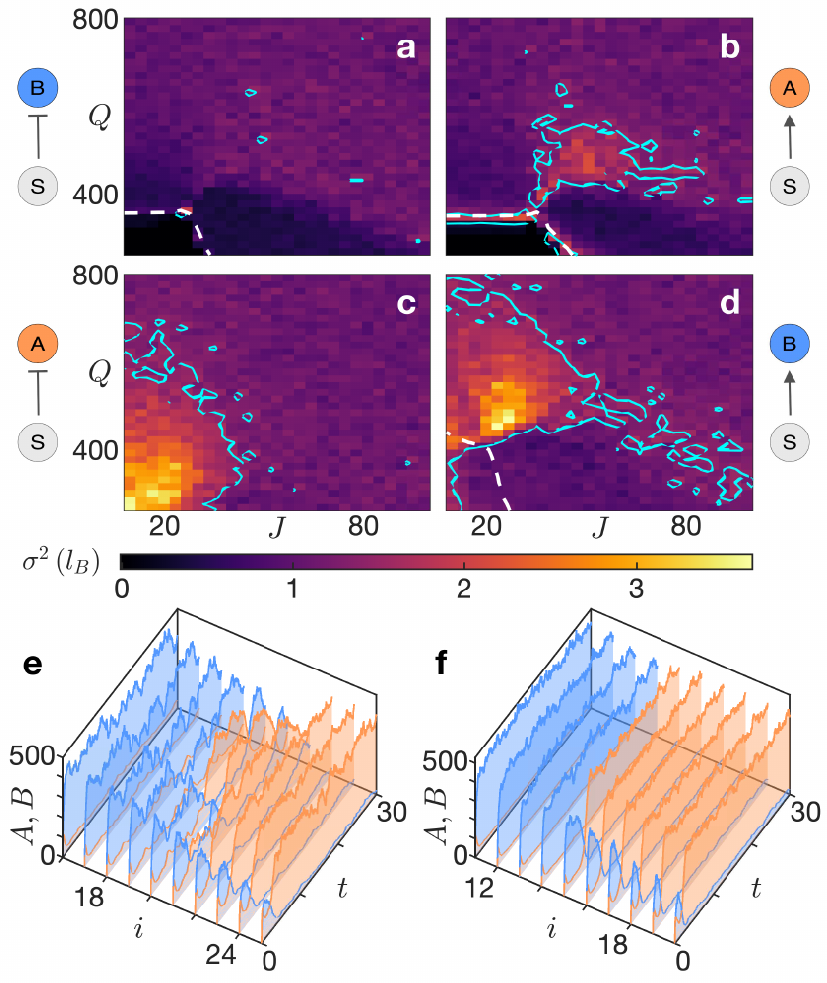}
\caption{{\bf Robust determination of cell fates results from interaction between stochastic
gene expression dynamics and contact-mediated signaling.}
The intercellular interactions mediated by the Notch downstream signal ($S$)
can be classified into four types, determined by which of the patterning
genes ($A$ or $B$) is either up or downregulated by $S$, as represented
by the motifs  shown beside each panel (a-d) [arrows representing up/downregulation
are as indicated in Fig.~\ref{fig:fig1}~(e)].
For each type, the spatial pattern formed by cells adopting distinct fates $A$, $B$
in a 1D domain comprising $N(=50)$ cells subject to a morphogen gradient
is characterized by the location $l_B$ of the boundary [$\sim 20$, in absence of any interactions between the cells, see Fig.~\ref{fig:fig1}~(c)] demarcating the segments expressing
the two fates. The variance in $l_B$ across $300$ stochastic realizations is shown for
each choice of the pair of parameters quantifying the strength of intercellular coupling, viz.,
$J$ representing critical value of patterning
gene expression segregating low/high receptor production and $Q$ representing
critical signal intensity that distinguishes between
weak and strong regulation of patterning gene expression.
The continuous curves in each panel are contours indicating the variance in $l_B$ in
the absence of intercellular interactions ($\simeq 1.38$).
The regions in the $J-Q$ plane above the broken curves (shown in white) correspond to the mean
value of $l_B$ lying within $[10,30]$, i.e., $50\%$ of its value in the uncoupled case.
Note that, for coupling types in which $S$ upregulates $A$ either directly (b), or indirectly via
suppression of its inhibitor $B$ (a), fluctuations in $l_B$  are markedly reduced
over a wider range of $J$ and $Q$.
(e-f) Temporal evolution of the expression of $A$ and $B$ shown for cells
around $l_B$ for the uncoupled case, contrasting (e) the
dynamics seen in absence of any intercellular interactions, with (f) that obtained when
$S$ inhibits $B$ [as in panel (a)]. While the uncoupled cells exhibit
large fluctuations in expression levels with uncertainty in $l_B$ sustained for a long time, in
the presence of intercellular interactions
cells rapidly converge to their eventual fates.}
\label{fig:fig2}
\end{figure}

To quantitatively characterize the role of inter-cellular interaction in
promoting robustness to noise,
we compare the variance of the spatial location of the fate boundary
when the cells interact via Notch signaling, with the case when the cells attain their fates
independent of their neighbors. The situation when the cells are uncoupled is shown in
Fig~\ref{fig:fig1}~(c), which displays the spatial distribution of steady state expression values of the
patterning genes. It is seen that in cells close to the fate boundary (i.e., $i \sim 20$)
the level of expression of both genes vary over a large range, with a substantial degree of
overlap between the two distributions.
As a result, the fates attained by each of these cells vary from one realization to another [Fig~\ref{fig:fig1}~(d), right],
which suggests that they have insufficient positional information for their eventual
identities to be determined with any certainty. This ambiguity in cell fates can lead to a
biologically undesirable outcome, viz., high variability in embryonic patterning
across individuals of a species.

The strength of the interaction between Notch signaling and patterning gene expression dynamics
is regulated in our model
by the parameters $Q$, $J$ and $K_5$ (see the expressions for $\mathcal{F}_{A,B}$, $\mathcal{F}_R$ and $\mathcal{F}_D$, respectively, defined above).
Two additional parameters $\gamma_A$ and $\gamma_B$ also play a role but only when
the signal $S$ \textit{upregulates} the patterning genes (see Table~\ref{tab:tab1}).
Here we focus on the two half-saturation constants $Q$ and $J$, where $Q$ is the
magnitude of $S$ above which the signal noticeably affects patterning gene expression, while $J$ determines the expression levels of the
patterning genes above which production of Notch receptors is appreciably increased.
The parameter $K_5$ which controls the strength of repression of the Delta ligand by
the Notch signal also contributes to the final outcome.
However, as the coupling-induced suppression of noise occurs even when $S$ has no effect
on $D$ production, we may conclude that the phenomenon is not critically dependent on
the value of $K_5$.

Fig.~\ref{fig:fig2} shows the dispersion in the fate boundary position
in a linear array of $N$ cells coupled via Notch-Delta signaling
as each of the parameters $Q$ and $J$ are varied over a large range,
for the four distinct types of inter-cellular interactions mentioned above.
While for all interactions we observe regions exhibiting a substantial reduction in the extent to which
the location $l_B$ of the fate boundary fluctuates across realizations, this is most prominent
when the interaction involves either $S$ downregulating the expression of $B$
[Fig.~\ref{fig:fig2}~(a)], or equivalently,
upregulating the expression of its inhibitor $A$ [Fig.~\ref{fig:fig2}~(b)].
We observe not only a much larger area of the $Q-J$ parameter space where the variance
$\sigma^2 (l_B)$
is lower than that for the case when inter-cellular interactions are absent, but also
a relatively greater certainty with which the domains exhibiting different fates are demarcated for these
two types of interactions.

\begin{figure}[tbp!]
\centering
\includegraphics[width=0.99\columnwidth]{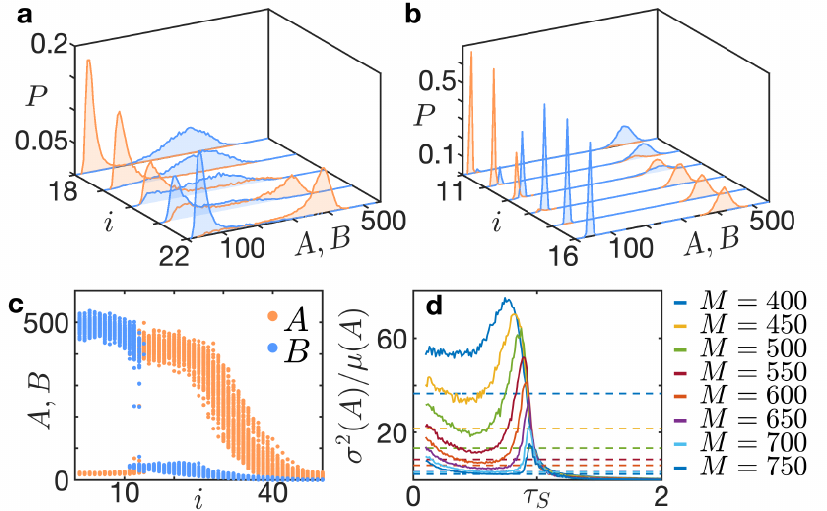}
\caption{\textbf{Reduction in variability of response to fluctuating morphogen concentrations
depends on relative time-scales of gene expression dynamics and contact-mediated signaling.}
(a-b) Steady-state distributions for the expression levels of the patterning genes $A$ and $B$
shown for cells located around the respective positions of the fate boundary when (a) intercellular interactions are absent, or (b) the Notch downstream signal $S$ suppresses expression of $B$
[as in Fig.~\ref{fig:fig2}~(a)].
In the uncoupled case, the distributions are extremely broad with a high degree of overlap
close to the fate boundary, indicating a large degree of uncertainty in the identity of the
gene having higher expression levels, and hence in the fate of the corresponding cells.
Intercellular interactions result in the gene expressions exhibiting sharply
defined peaks at either very low or very high levels, with the gene that is dominantly expressed
at any given cell clearly identifiable. This leads to a steady state expression of the patterning genes
[shown in (c) for a 1-dimensional array comprising $50$ cells] that exhibits a robust,
sharply defined
cell fate boundary (at $i \approx 12$) even in the presence of a noisy morphogen gradient.
Results of $300$ different realizations are shown.
(d) Temporal variance in the expression of gene $A$ in a given cell, expressed relative to its
mean value, shown as a function of the mean lifetime $\tau_S$ of the Notch downstream
signal $S$. For different mean concentrations $M$
(indicated by distinct colors, see legend) of the morphogen,
a peak is observed at a critical value of $\tau_S$ above which the system is effectively
insensitive to fluctuations. For each $M$, a broken horizontal line (of the same color) represents the corresponding variance:mean ratio for the uncoupled case, i.e., in the absence of the Notch signal.
}
\label{fig:fig3}
\end{figure}

This enhanced robustness of the cell fate pattern when $S$ suppresses $B$ (or equivalently,
promotes $A$) can be understood in terms of the
alteration in the steady-state level of expression of the patterning genes around the
fate boundary.
In the absence of coupling, not only are the expression levels of both genes distributed
over a larger range for each cell, but the two distributions also exhibit a substantial
degree of overlap [Fig.~\ref{fig:fig3}~(a)]. This suggests that the identity of the gene which
eventually dominates at the steady state (and hence decides the fate)
for any cell close to the boundary is largely decided by stochastic perturbations.
In contrast, the inter-cellular interactions result in suppression of the patterning
gene $B$ by $S$ specifically in the region of the array where the two patterning genes are
expressed at comparably high levels ($>J$) and consequently, where the two distributions overlap.
Thus, we observe from Fig.~\ref{fig:fig3}~(b) that for cells ($i \geq 13$) where both $A,B >J$
in the steady state for the uncoupled case, the Notch-mediated interaction leads to the dominance
of $A$ over $B$ consistently across all realizations. The repression of $B$ by $A$ results in
the peaks of their respective distributions becoming widely separated. For cells closer to the
morphogen source ($i \leq 12$), $B$ dominates because of the asymmetric strength
of mutual repression between the two patterning genes mentioned earlier,
resulting in low expression levels of $A$ and
consequently, negligible production of $S$. Hence, for these cells also we observe widely
separated peaks for $A$ and $B$ distributions, but with the latter occurring at higher
values (as in the uncoupled case). The inter-cellular interactions can, thus, be seen
as enhancing the distinction between the steady-state levels of $A$ and $B$, the elimination
of overlap between the two distributions leading to a sharply defined fate boundary
[Fig.~\ref{fig:fig3}~(c), compare with Fig.~\ref{fig:fig1}~(c)].
Note that boundary shifts closer to the morphogen source (with respect to its location in the uncoupled case), as the inter-cellular interactions in which $S$ suppresses $B$ (or promotes $A$) favors the dominance of $A$ where the two overlap in the absence of interactions.
Our model, thus, helps explain the shift in fate boundary that has been observed when
cells communicate via Notch signaling~\cite{Kong2015,Kuyyamudi2021}.
Consistent with this explanation, the reverse is observed for types of interaction where $S$ instead
suppresses $A$
(or promotes $B$) with the fate boundary location moving further away from the morphogen
source [see Supplementary Information].

The mechanism of interaction between $S$ and the patterning genes can be made more
transparent by considering a simplified scenario where the \textit{trans} ligand concentration
in the neighborhood that stimulates the receptors of a cell is assumed to be time-invariant.
Such an approximation is still capable of reproducing the phenomenon of noise-suppression,
which is not crucially dependent on the dynamics of $D$.
We investigate the patterning gene expression dynamics in the cells of such a system, subjected
to stochastic fluctuations in morphogen concentration (around the mean value $M$) and intrinsic noise.
Fig.~\ref{fig:fig3}~(d) shows the relative variance in the expression of gene $A$ ($B$ exhibits
qualitatively similar behavior, see Supplementary Information) as a function of the mean lifetime $\tau_S$ of the
downstream effector for the Notch signaling pathway. We observe that independent of the
mean morphogen concentration (and hence, the position of a cell on an array that is subject to
a morphogen gradient),
the gene expression level
becomes extremely robust to noise when $\tau_S$ is sufficiently large ($\gtrsim1$).
To understand this, we note from the expression for $\mathcal{F}_S$ (see equation above)
that increasing $\tau_S$ results in a proportionately higher steady state value of $S$ that
a cell is subjected to. Focusing on the interaction in which $S$ downregulates $B$ expression,
we note that for high values of $S$ the dynamics of $B$ is altered as
the function $\Phi_B$ essentially reduces to zero (for reasonably high values of the Hill exponent $g$).
The resultant sharp decrease in the production terms in $\mathcal{F}_B$ implies that
$A$ will dominate $B$ in all cells where $S$ is high. As the magnitude of the signal also
depends on receptor concentration, whose production is high only for those cells in which
both $A$ and $B$ are expressed at sufficiently high levels ($>J$), the $S$-induced suppression
of $B$ will only be observed in those cells where the distributions of the patterning genes
overlap considerably.
Similar behavior will be seen for the interaction where $S$ upregulates expression of $A$,
as the latter inhibits $B$ leading to effective downregulation of $B$ by $S$.

For the type of interaction in which the signal downregulates $A$ (or equivalently, upregulates $B$),
the function $\Phi_A$, and hence the production term in $\mathcal{F}_A$, decreases to very low values for large $S$. As a result, $B$ is favored to dominate over $A$ in the region where
the patterning genes are expressed at comparable levels when the cells are uncoupled.
This would lead one to expect an analogous situation to that described above but with
$B$ replacing $A$ as the preferred cell fate around the fate boundary location for the non-interacting case. However, as this region is located
relatively far from the morphogen source, the local concentration of $M$ may not be
high enough to promote the expression of $B$ while being sufficient for the expression
of $A$ (as $K_2 > K_1$). As a result, the advantage conferred to $B$ by the contact-mediated
interaction is offset by the low morphogen concentration that favors $A$, preventing outright
dominance by either gene in this region. Hence, these two types of interactions between
$S$ and the patterning genes are unable to reduce the variability in fate boundary position
for a wide range of choices of the parameters $Q$ and $P$ [Fig.~\ref{fig:fig2}~(c-d)].

To conclude, we have shown that contact-mediated interaction between cells
(e.g., via downstream signaling triggered by binding of Notch receptors on a cell surface with
the surface-bound ligands of its neighbors) can reduce the uncertainty in cell fates
that arise from stochastic fluctuations in the morphogen concentration that
provides positional information to the cells, as well as, intrinsic noise. Even though the
signaling mechanism we employ is also subject to random variability in its components,
the coupling between cells that it effects is able to markedly reduce the dispersion in
the position of the boundary between regions expressing distinct cell fates and thus
enhancing robustness of spatial patterns arising in tissues and organs over the course of development. Our results suggest a functional role for the higher level of Notch
signaling observed in the cells demarcating the boundary between the regions expressing
dorsal and ventral fates in the \textit{Drosophila} hindgut~\cite{Fuss2002}. Notch activity
is also known to be crucial for defining the boundaries of the organ of corti in the cochlea
of mice, consistent with the mechanism outlined here~\cite{Basch2016}. A more
direct experimental test of our model can involve verifying that those regions in tissue undergoing
differentiation, whose cells
have comparable levels of expression for the different patterning genes, exhibit higher
levels of Notch activity.
The results reported here show that the nature of interaction between the downstream effector
of the intercellular signaling mechanism and the patterning gene(s) is important in
determining the extent to which coupling between cells enhance the robustness of
cell fate patterns. In particular, they suggest that the mechanism is more effective
in suppressing noise and reducing variability when Notch signaling upregulates
that patterning gene (or equivalently, downregulates the gene repressing it) which requires a relatively lower concentration of the morphogen to be expressed.
This is a potential experimental test for the proposed model, involving comparison of
expression levels of different patterning genes in the presence of inter-cellular
interactions with that observed in its absence (e.g., implemented by knocking out Notch).


We would like to thank Marcin Zag\'{o}rski for helpful discussions.
SNM has been supported by the IMSc Complex Systems Project (12th
Plan), and the Center of Excellence in Complex Systems and Data
Science, both funded by the Department of Atomic Energy, Government of
India. The simulations required for this work were
supported by IMSc High Performance Computing facility (hpc.imsc.res.in) [Nandadevi].

%

\clearpage
\onecolumngrid

\setcounter{figure}{0}
\renewcommand\thefigure{S\arabic{figure}}
\renewcommand\thetable{S\arabic{table}}

\vspace{1cm}
\begin{center}
\textbf{\large{SUPPLEMENTARY INFORMATION}}\\

\vspace{0.5cm}
\textbf{\large{Precision of morphogen-driven tissue patterning during development is enhanced through contact-mediated cellular interactions}}\\
\vspace{0.5cm}
\textbf{Chandrashekar Kuyyamudi, Shakti N. Menon and Sitabhra Sinha}
\end{center}
\section*{List of Supplementary Figures}
\begin{enumerate}
\item Fig S1: Precision of cell fate determination resulting from interaction
between stochastic gene expression dynamics and contact-mediated signaling measured
in terms of steepness of spatial profile for expression of gene~$A$.
  \item Fig S2: Precision of cell fate determination resulting from interaction
between stochastic gene expression dynamics and contact-mediated signaling measured
in terms of steepness of spatial profile for expression of gene~$B$.
  \item Fig S3: The location $l_B$ of the fate boundary in a linear array of cells resulting from
different types of interaction
between stochastic gene expression dynamics and contact-mediated signaling, in the
presence of a morphogen gradient.
\end{enumerate}

\begin{figure}[ht]
\includegraphics[width=\textwidth]{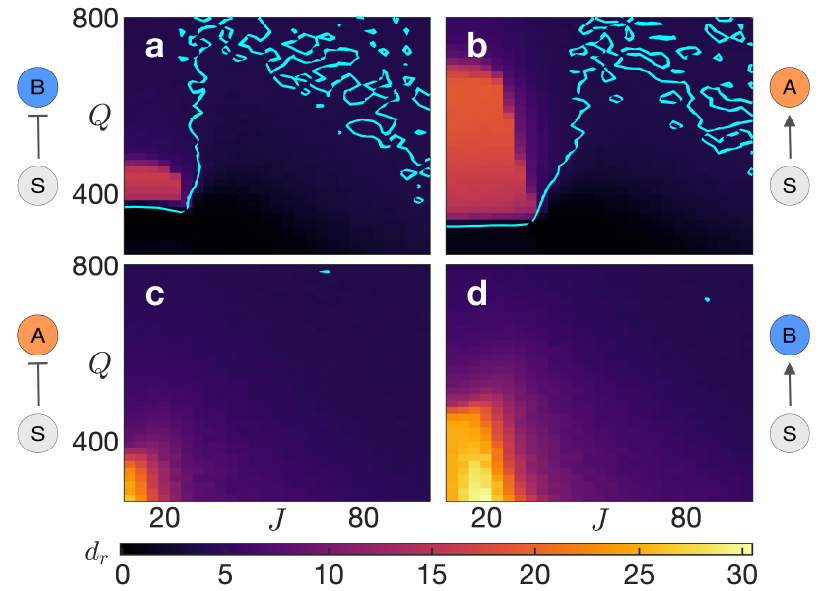}
\caption{{\bf Precision of cell fate determination resulting from interaction
between stochastic gene expression dynamics and contact-mediated signaling measured
in terms of steepness of spatial profile for expression of gene $A$.} 
For each of the four types of interaction between the Notch downstream signal $S$ and the
patterning genes $A,B$ (represented by the motifs shown beside each panel), the precision of
the spatial pattern formed by cells adopting distinct fates $A$, $B$
in a 1D domain comprising $N(=50)$ cells subject to a morphogen gradient
is characterized by the slope of the growth in steady state expression levels of $A$ across the cell array 
[the corresponding curve is shown in Fig.~1~(c) in the main text for the case when 
interactions are absent between cells]. 
This is measured by the rise distance, viz., the width (measured in terms of number of cells)
over which $A$ change from $10\%$ to $80\%$ of its maximum expression value for the
type of interaction being considered. 
The mean rise distance ($d_r$) across $300$ stochastic realizations is shown for
each choice of the pair of parameters quantifying the strength of intercellular coupling, viz.,
$J$ representing critical value of patterning
gene expression segregating low/high receptor production and $Q$ representing
critical signal intensity that distinguishes between
weak and strong regulation of patterning gene expression.
The continuous curves in each panel are contours indicating the variance in $d_r$ in
the absence of intercellular interactions ($\simeq 3.78$).
Note that, for coupling types in which $S$ upregulate $A$ either directly (b), or indirectly via
suppression of its inhibitor $B$ (a), intercellular interactions are able to markedly
increase the steepness of the spatial profile of gene expression, resulting in 
a sharply defined fate boundary, over a wider range of coupling strengths $J$ and $Q$.
In contrast, the resolution achieved with coupling types in which $S$ upregulates $B$ 
either directly (d), or indirectly via suppression of its inhibitor $A$ (c), is almost always
lower than even the uncoupled case.}
\label{figS1}
\end{figure}

\begin{figure}[ht]
\includegraphics[width=\textwidth]{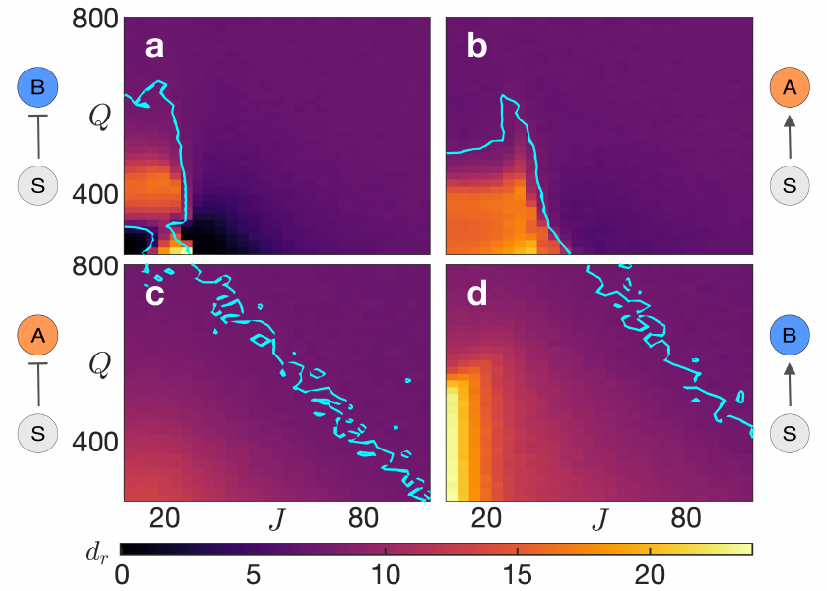}
\caption{{\bf Precision of cell fate determination resulting from interaction
between stochastic gene expression dynamics and contact-mediated signaling measured
in terms of steepness of spatial profile for expression of gene $B$.}
For each of the four types of interaction between the Notch downstream signal $S$ and the
patterning genes $A,B$ (represented by the motifs shown beside each panel), the precision
of the spatial pattern formed by cells adopting distinct fates $A$, $B$
in a 1D domain comprising $N(=50)$ cells subject to a morphogen gradient
is characterized by the slope of the decline in steady state expression levels of $B$ 
across the cell array 
[the corresponding curve is shown in Fig.~1~(c) in the main text for the case when 
interactions are absent between cells]. 
This is measured by the rise distance, viz., the width (measured in terms of number of cells)
over which $B$ change from $80\%$ to $10\%$ of its maximum expression value for the
type of interaction being considered. 
The mean rise distance ($d_r$) across $300$ stochastic realizations is shown for
each choice of the pair of parameters quantifying the strength of intercellular coupling, viz.,
$J$ representing critical value of patterning
gene expression segregating low/high receptor production and $Q$ representing
critical signal intensity that distinguishes between
weak and strong regulation of patterning gene expression.
The continuous curves in each panel are contours indicating the variance in $d_r$ in
the absence of intercellular interactions ($\simeq 7.47$).
Note that, for coupling types in which $S$ upregulate $A$ either directly (b), or indirectly via
suppression of its inhibitor $B$ (a), intercellular interactions are able to markedly
increase the steepness of the spatial profile of gene expression, resulting in 
a sharply defined fate boundary, over a wider range of coupling strengths $J$ and $Q$.
The region of ($J,Q$) parameter space over which a higher resolution than the uncoupled
case can be achieved is much reduced for the coupling types in which $S$ upregulates $B$ 
either directly (d), or indirectly via suppression of its inhibitor $A$ (c).}
\label{figS2}
\end{figure}

\begin{figure}[ht]
\includegraphics[width=\textwidth]{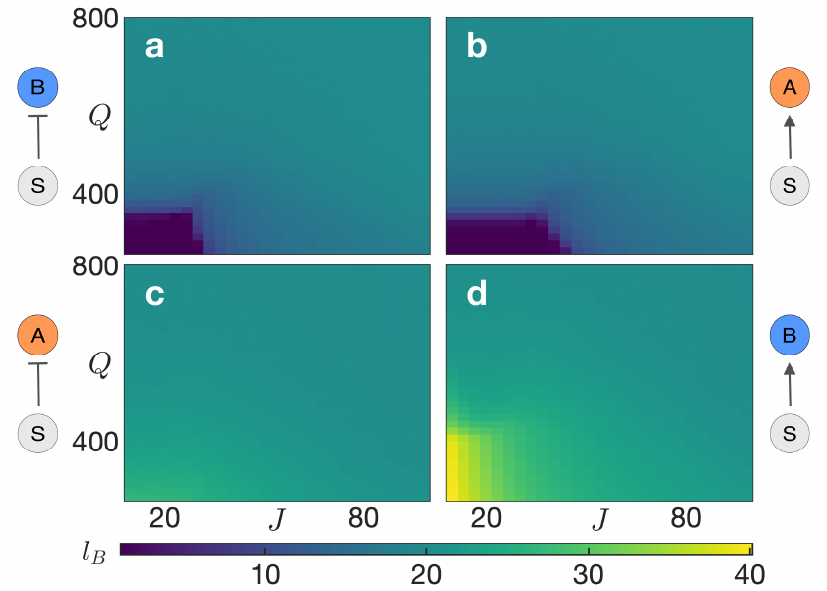}
\caption{{\bf The location $l_B$ of the fate boundary in a linear array of cells resulting from
different types of interaction
between stochastic gene expression dynamics and contact-mediated signaling, in the
presence of a morphogen gradient.}
For each of the four types of interaction between the Notch downstream signal $S$ and the
patterning genes $A,B$ (represented by the motifs shown beside each panel), the spatial pattern formed by cells adopting distinct fates $A$, $B$
in a 1D domain comprising $N(=50)$ cells subject to a morphogen gradient
is characterized by the location $l_B$ of the boundary [$\sim 20$, in absence of any interactions between the cells, see Fig.~1~(c) in main text] demarcating the segments expressing
the two fates. The mean position of the boundary across $300$ stochastic realizations is shown for
each choice of the pair of parameters quantifying the strength of intercellular coupling, viz.,
$J$ representing critical value of patterning
gene expression segregating low/high receptor production and $Q$ representing
critical signal intensity that distinguishes between
weak and strong regulation of patterning gene expression.
Note that, for coupling types in which $S$ upregulate $A$ either directly (b), or indirectly via
suppression of its inhibitor $B$ (a), intercellular interactions result in the fate boundary
moves towards the morphogen source (in comparison to the uncoupled case). In contrast,
the fate boundary moves further away from the morphogen source for coupling types in 
which $S$ upregulates $B$ either directly (d), or indirectly via suppression of its inhibitor $A$ (c).
}
\label{figS3}
\end{figure}

\end{document}